\documentclass[conference]{IEEEtran}
\IEEEoverridecommandlockouts
\usepackage{cite}
\usepackage{amsmath,amssymb,amsfonts}
\usepackage{algorithm}
\usepackage[noend]{algpseudocode}
\usepackage{graphicx}
\usepackage{textcomp}
\usepackage{xcolor}
\usepackage{acronym}
\usepackage{multirow}
\usepackage{amssymb}
\usepackage{pifont}
\usepackage{todonotes}
\usepackage{colortbl}

\newcommand{\sidepubstatus}[1]{%
  \begin{tikzpicture}[remember picture,overlay]
    \node[rotate=90,anchor=center,inner sep=0pt,font=\small] 
      at ([xshift=-0.5cm]current page.east |- current page.center) {#1};
  \end{tikzpicture}%
}

\def\BibTeX{{\rm B\kern-.05em{\sc i\kern-.025em b}\kern-.08em
    T\kern-.1667em\lower.7ex\hbox{E}\kern-.125emX}}

\definecolor{ceil}{rgb}{0.57, 0.63, 0.81}

\begin{document}

\acrodef{api}[API]{Application Programming Interface}
\acrodef{ai}[AI]{Artificial Intelligence}
\acrodef{cpu}[CPU]{CPU}
\acrodef{gpu}[GPU]{Graphics Processing Unit}
\acrodef{npu}[NPU]{Neural Processing Unit}
\acrodef{xpu}[XPU]{Custom Acceleration Processing Unit}
\acrodef{dvfs}[DVFS]{Dynamic Voltage/Frequency Scaling}
\acrodef{hmp}[HMP]{Heterogeneous Multi-Processing}
\acrodef{hsa}[HSA]{Heterogeneous System Architecture}
\acrodef{os}[OS]{Operating System}
\acrodef{tdp}[TDP]{Thermal Design Power}
\acrodef{gn}[GN]{Gateway Node}
\acrodef{ln}[LN]{Local Node}
\acrodef{iot}[IoT]{Internet of Things}
\acrodef{ml}[ML]{Machine Learning}
\acrodef{qos}[QoS]{Quality of Service}
\acrodef{ftp}[FTP]{Fused Tile Partitioning}
\acrodef{aofl}[AOFL]{Adaptive Optimal Fused-layer}
\acrodef{socs}[SoCs] {System-on-Chips}
\acrodef{rtm}[RTM]{Run-time Resource Management}
\acrodef{dnn}[DNN]{Deep Neural Network}
\acrodef{dop}[DoP]{Degree of Parallelism}
\acrodef{ppw}[PPW]{Performance per Watt}
\acrodef{os}[OS]{Operating System}
\acrodef{fsm}[FSM]{Finite State Machine}
\acrodef{lan}[LAN]{Local Area Network}
\acrodef{wlan}[WLAN]{Wireless Local Area Network}
\acrodef{elan}[ELAN]{Ethernet Local Area Network}
\acrodef{ilp}[ILP]{Integer Linear Programming}
\acrodef{eeg}[EEG]{Electroenceophelogram}
\acrodef{rl}[RL]{Reinforcement Learning}
\acrodef{bodp}[BODP]{Biased One-Dimensional Partition}
\acrodef{dl}[DL]{Deep Learning}
\acrodef{morl}[MORL]{Multi-Objective Reinforcement Learning}
\acrodef{momdp}[MOMDP]{Multi-Objective Markov Decision Process}
\acrodef{mdp}[MDP]{Markov's Decision Process}
\acrodef{Sarsa}[Sarsa]{State-Action-Reward-State’-Action’}
\acrodef{onnx}[ONNX]{Open Neural Network Exchange}
\acrodef{ar}[AR]{Augmented Reality}
\acrodef{vr}[VR]{Virtual Reality}
\acrodef{mr}[MR]{Mixed Reality}
\acrodef{dag}[DAG]{Directed Acyclic Graph}
\acrodef{mac}[MAC]{Multiply–Accumulate operation}
\acrodef{dse}[DSE]{Design Space Exploration}
\acrodef{ppo}[PPO]{Proximal Policy Optimization}
\acrodef{soa}[SoA]{State-of-the-Art}
\acrodef{cnn}[CNN]{Convolution Neural Network}
\acrodef{fcfs}[FCSFS]{First-come-first-serve}
\acrodef{wdg}[WDG]{Weighted Directed Graph}
\acrodef{dp}[DP]{Dynamic Programming}

\title{HiDP: Hierarchical DNN Partitioning for Distributed Inference on Heterogeneous Edge Platforms
\thanks{\textbf{This manuscript is accepted at the 28th Design, Automation and Test in Europe Conference (IEEE DATE, 2025). We gratefully acknowledge funding from EU Horizon 2020 Research and Innovation Programme under the Marie Sk\l{}odowska Curie grant No. $956090$, Approximate Computing for Power and Energy Optimisation (APROPOS).}}
}



\author{\IEEEauthorblockN{Zain Taufique}
\IEEEauthorblockA{
\textit{University of Turku}\\
Turku, Finland \\
zatauf@utu.fi}
\and
\IEEEauthorblockN{Aman Vyas}
\IEEEauthorblockA{
\textit{University of Turku}\\
Turku, Finland \\
amvyas@utu.fi}
\and
\IEEEauthorblockN{Antonio Miele}
\IEEEauthorblockA{
\textit{Politecnico di Milano}\\
Milan, Italy \\
antonio.miele@polimi.it}
\and
\IEEEauthorblockN{Pasi Liljeberg}
\IEEEauthorblockA{
\textit{University of Turku}\\
Turku, Finland \\
pasi.liljeberg@utu.fi}
\and
\IEEEauthorblockN{Anil Kanduri}
\IEEEauthorblockA{
\textit{University of Turku}\\
Turku, Finland \\
spakan@utu.fi}
}

\maketitle

\sidepubstatus{Accepted to be published in 28th Design, Automation and Test in Europe Conference (IEEE DATE, 2025)}

\begin{abstract}

Edge inference techniques partition and distribute Deep Neural Network (DNN) inference tasks among multiple edge nodes for low latency inference, without considering the core-level heterogeneity of edge nodes. Further, default DNN inference frameworks also do not fully utilize the resources of heterogeneous edge nodes, resulting in higher inference latency. In this work, we propose a hierarchical DNN partitioning strategy (\textit{HiDP}) for distributed inference on heterogeneous edge nodes. Our strategy hierarchically partitions DNN workloads at both global 
and local levels by considering the core-level heterogeneity of edge nodes. We evaluated our proposed \textit{HiDP} strategy against relevant distributed inference techniques over widely used DNN models on commercial edge devices. On average our strategy achieved 38\% lower latency, 46\% lower energy, and 56\% higher throughput in comparison with other relevant approaches.

\end{abstract}


\acused{gpu}
\acused{cpu}

\section{Introduction}
\acp{dnn} enable a wide range of applications such as augmented reality, smart glasses, and live video analytics, etc~\cite{band}. Such applications demand real-time low latency \ac{dnn} inference over continuous streaming input data \cite{dnn_mapping_scenario_based}. Offloading \ac{dnn} inference to remote cloud servers can lead to unpredictable latency with communication penalty, while local edge devices have limited compute capabilities to provide low latency inference~\cite{DeepThings}. Edge inference techniques distribute the inference workload among a cluster of collocated edge nodes to improve \ac{dnn} inference latency \cite{MoDNN,DeepThings,disnet}. 

Existing distributed edge inference strategies partition the inference workload into \textit{blocks} by splitting the layers of a \ac{dnn} model \cite{isplit, omniboost, partner, roadrunner, pipeit, moc, hasp} and/or the input data of the \ac{dnn} inference request \cite{MoDNN, DeepThings, deepslicing, Legion, AutoDice, eDNN}. Subsequently, these \textit{blocks} are distributed to different edge nodes within the edge cluster, based on the compute capacity of an edge node \cite{DeepThings,MoDNN,AspDac} and the computation-communication ratio of a \textit{block}~\cite{disnet,dnn_mapping_scenario_based,omniboost}. It should be noted that edge nodes are composed of heterogeneous processing units including CPU, \ac{gpu}, and \acp{npu}, exhibiting a high degree of compute diversity within the single node and across the edge cluster. However, existing distributed inference techniques make global decisions on \ac{dnn} partitioning and distribution, without considering the core-level heterogeneity of individual edge nodes. 

After the creation and distribution of \textit{blocks}, aforementioned distributed inference techniques rely on deep learning frameworks to schedule the inference service on a local edge device. Deep learning frameworks do not fully utilize the resources of a heterogeneous multi-core edge node, leading to sub-optimal inference latency. For example, TensorFlow~\cite{tensorflow} schedules inference on GPU by default, unless explicitly specified by the application to use other CPUs/NPUs \cite{gpu_placement_tf, gpu_placement_tf_princeton}. Running inference on a single processing unit (e.g. GPU) misrepresents the compute capacity of the local heterogeneous edge node, resulting in sub-optimal workload partitioning and distribution decisions made on a global level. This problem is emphasized on edge platforms with CPUs performing better than GPUs \cite{AutoScale} \cite{pipeit}, and while running CPU-friendly layers of \ac{dnn} models \cite{arm-co-up}. Advanced inference strategies that consider core-level heterogeneity are tailored for inference on a single edge node \cite{pipeit} \cite{arm-co-up} \cite{band}. Adapting heterogeneity-aware \ac{dnn} scheduling techniques for distributed inference requires intelligent orchestration to jointly optimize \ac{dnn} partitioning and workload distribution, along with infrastructural support for inter-node data and decision control transfer. However, existing distributed inference strategies lack such intelligent orchestration.

\begin{figure}
    \centering
    \includegraphics[width=0.489\textwidth]{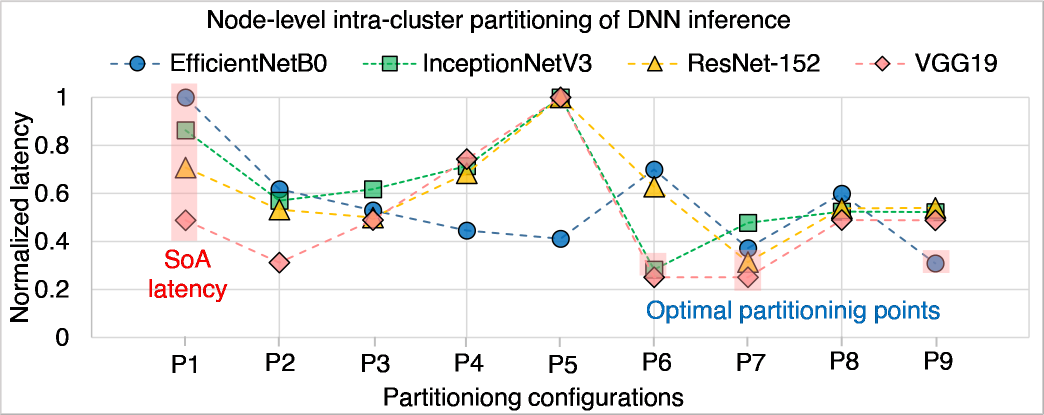}
    \vspace{-12pt}
    \caption{Inference latency of DNN models with different workload partitioning configurations.}
    \vspace{-15pt}
    \label{fig.inter-cluster}
\end{figure}


We demonstrate the limitations of existing distributed inference techniques over four \ac{dnn} models run on the Jetson TX2 platform \cite{nvidia_jetson_tx2}. Figure \ref{fig.inter-cluster} shows normalized inference latency of the \ac{dnn} models with different workload partitioning configurations (\texttt{P1-P9}). Each partitioning configuration represents a specific combination of (i) number of data-wise partitions of the \ac{dnn} model and (ii) CPU-GPU workload split. Among these configurations, P1 is the default TensorFlow workload scheduling choice -- running inference exclusively on the GPU with no data partitioning. State-of-the-art distributed inference techniques use the workload partitioning configuration P1 (highlighted as \textit{SoA latency} in Figure \ref{fig.inter-cluster}) on a local edge device, resulting in higher inference latency. However, inference latencies of all the \ac{dnn} models are lower in partitioning configurations other than P1, where the inference workload is partitioned and split between the CPU and GPU. Latency of ResNet-152 and VGG-19 are lowest at P7 (4 data partitions with 80\% workload on GPU and 20\% on CPU), InceptionNet-V3 at P6 (90\% workload with 2 data partitions on GPU and 10\% workload with 4 data partitions on CPU), and EfficientNet-B0 at P9 (4 data partitions and 50\% workload split between CPU and GPU). InceptionNet-V3, ResNet-152, VGG-19, and EfficientNet-B0 have 65\%, 40\%, 25\%, and 75\% lower latency respectively, in comparison with existing distributed inference strategies using the default TensorFlow run-time. It should be noted that optimal partitioning configuration (number of partitions and workload split between CPU and GPU) differs for different \ac{dnn} models. Our analysis highlights that the default partitioning configuration used by existing distributed inference techniques -- (i) results in higher inference latency on local edge devices, and (ii) skews workload partitioning decisions exclusively made at a global level. These limitations can be addressed by considering the core-level heterogeneity of edge nodes while making both global and local workload partitioning decisions. 

In this work, we propose a two-tier hierarchical \ac{dnn} partitioning (\textit{HiDP}) strategy to improve the latency of distributed inference over a cluster of heterogeneous edge devices. Our approach considers both core-level and device-level heterogeneity among edge devices within the cluster to combinatorially determine global \ac{dnn} partitioning and workload assignment to edge nodes, followed by local \ac{dnn} partitioning. To the best of our knowledge, ours is the first work that considers hybrid partitioning for distributed \ac{dnn} inference in heterogeneous edge nodes while optimizing core-level resource usage. Our novel contributions include:
\begin{itemize}
    \item Hierarchical \ac{dnn} partitioning strategy for minimizing latency of distributed inference on edge platforms
    \item Collaborative edge cluster framework for partitioning, distribution, and scheduling of \ac{dnn} inference services 
    \item Evaluation of HiDP strategy against existing distributed inference techniques on real hardware edge cluster setup including Jetson Orin NX, Jetson Nano, Jetson TX2, Raspberry Pi 4B, and Raspberry Pi 5.
\end{itemize}




The paper is organized as follows: Section~\ref{sec.mot} provides background and motivation for our proposed approach, and Section ~\ref{sec.frm} presents an overview of our framework infrastructure. Section~\ref{sec.exp} evaluates our proposed solution against other relevant strategies, followed by conclusions in Section~\ref{sec.conc}.

\section{Background and Motivation} \label{sec.mot}

\subsection{DNN partitioning}

For distributed inference, \acp{dnn} can be partitioned model-wise \cite{MoDNN} and data-wise \cite{eDNN}.
In \textit{model partitioning}, \ac{dnn} layers are dynamically grouped into executable \textit{blocks}; these blocks are offloaded to multiple devices for distributed execution in a pipelined fashion. This strategy is inherently temporal since layers across different \ac{dnn} \textit{blocks} are executed sequentially. Minimizing inference latency requires creation of variable-sized \ac{dnn} \textit{blocks} dynamically by considering resources across the edge cluster. Model partitioning is feasible for dense \ac{dnn} models that enable coarse creation of compute-intensive \textit{blocks}. 
Alternatively, \textit{data partitioning} splits the input data and creates smaller-sized sub-models of the original model for parallel execution.
Data partitioned inference is spatio-temporal, since parallelly executed sub-models exchange intermediate data to maintain accuracy. Workloads with larger input sizes are suitable for data partitioning, considering the computation-communication ratio of intermediate data sharing. Minimizing inference latency of a \ac{dnn} model on edge clusters requires joint selection of feasible partitioning strategy, optimal partitioning points, and workload distribution and scheduling.

\begin{figure}
    \centering
    \vspace{-3pt}
    \includegraphics[width=0.49\textwidth]{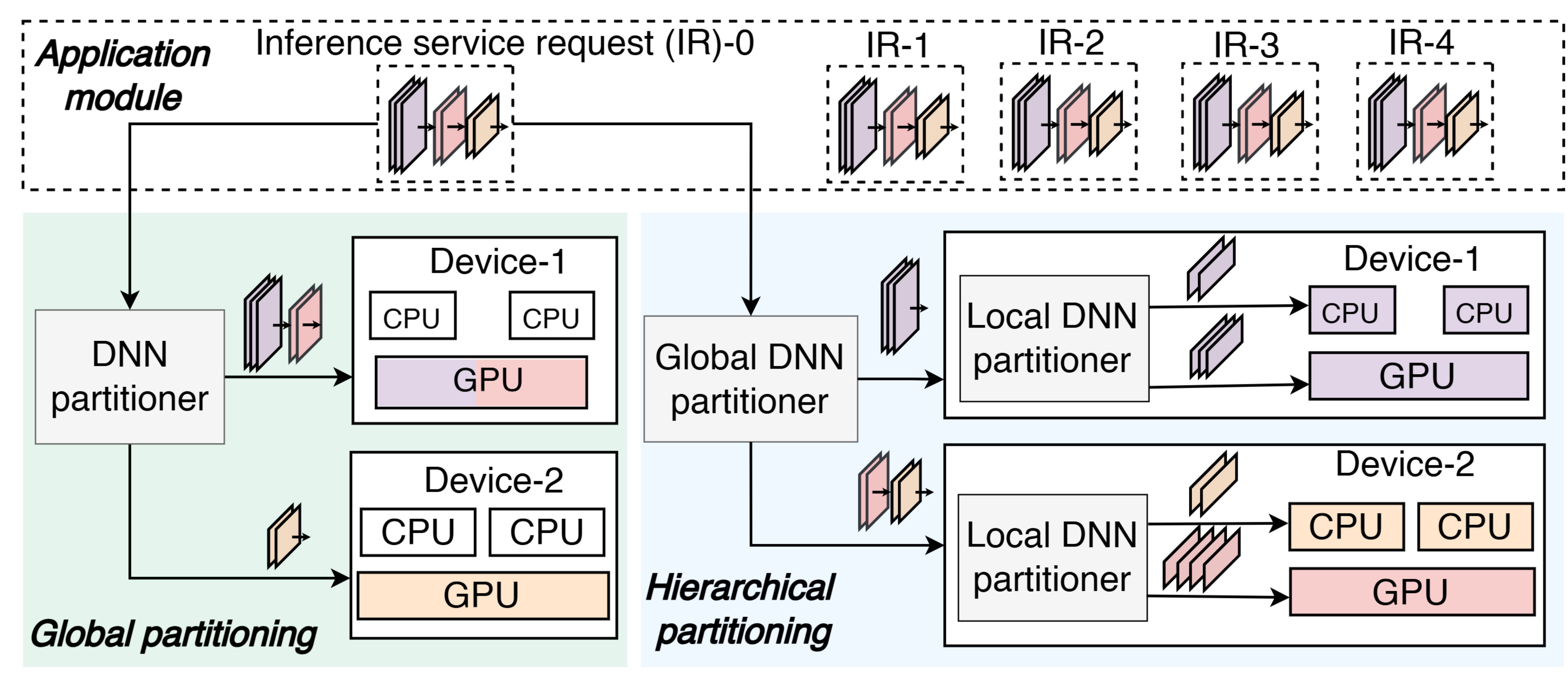}
    \vspace{-15pt}
    \caption{Comparison of global and hierarchical \ac{dnn} partitioning strategies.}
    \vspace{-15pt}
    \label{fig.intro}
\end{figure}

\begin{figure*}[t]
    \centering
    \vspace{-3pt}
    \includegraphics[width=0.85\textwidth]{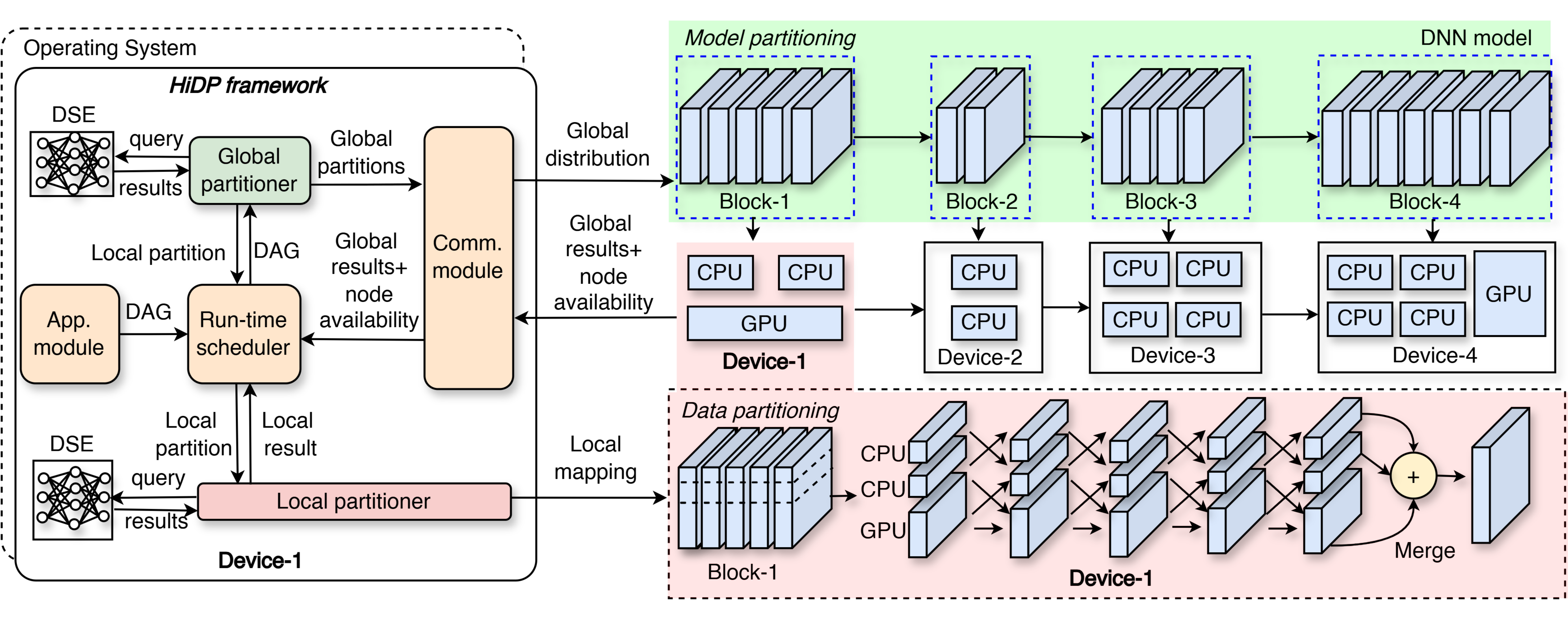}
    \vspace{-12pt}
    \caption{Overview of the proposed \textit{HiDP} framework. In this instance, the framework is run on device-1, partitioning the \ac{dnn} model globally through model partitioning and locally through data partitioning.}
    \vspace{-15pt}
    \label{fig:sys}
\end{figure*}
\vspace{-3pt}
\subsection{Motivational Example}
We demonstrate the benefits of hierarchical \ac{dnn} partitioning in comparison with global partitioning strategy over a motivational scenario. Figure \ref{fig.intro} shows an exemplar distributed \ac{dnn} inference scenario, with a sequence of \ac{dnn} inference requests (IR0-IR4) to be executed on two heterogeneous edge nodes, device-1 and device-2. With the global partitioning strategy (shown in green in Figure \ref{fig.intro}), the \textit{\ac{dnn} Partitioner} creates layer-wise blocks and distributes them among devices 1 and 2. Within devices 1 and 2, the inference blocks are executed on GPU by default. As presented empirically in Figure \ref{fig.inter-cluster}, this partitioning configuration leads to longer inference latency, potentially affecting other \ac{dnn} inference requests in the queue. With the hierarchical partitioning strategy (shown in blue in Figure \ref{fig.intro}), the \textit{Global \ac{dnn} Partitioner} creates layer-wise blocks based on the core-level heterogeneity of devices 1 and 2. It should be noted that the hierarchical partitioning strategy creates blocks that are different from the global partitioning strategy. Subsequently, the \textit{Local \ac{dnn} Partitioner} on each device further partitions the \ac{dnn} blocks and allocates to all the available cores. Thus, the hierarchical partitioning strategy results in lower inference latency, while also freeing up resources for subsequent inference requests in the queue. This example scenario highlights the need for making heterogeneity-aware \ac{dnn} partitioning decisions globally, followed by local optimization for minimizing the inference latency.  


\subsection{Related Work}\label{sec.related}
Existing distributed inference techniques use \ac{dnn} model partitioning to schedule inference workload over multiple edge nodes in a pipelined fashion \cite{pipeit, band, moc, hasp} or offload the inference service to resourceful cloud \cite{isplit, omniboost, partner, roadrunner}. Pipelining the model partitioned \ac{dnn} inference is sequential and is beneficial for dense \ac{dnn} models with a continuous stream of inference requests. Other techniques focus on data partitioning by splitting the input data into batches \cite{AspDac}, sub-images \cite{MoDNN, eDNN, AutoDice}, or intermediate layers through \ac{ftp} \cite{DeepThings, Legion}. Data partitioning allows parallel and distributed inference. However, the communication overhead of intermediate data exchange becomes significantly high for \acp{dnn} with a smaller input size. Recently, DisNet \cite{disnet} proposed hybrid partitioning for \ac{dnn} inference to minimize the overheads of both partitioning techniques without considering granular control over local device resources. \textit{HiDP} performs hybrid partitioning using model and data partitioning while having fine-grained control over local edge node's resources to minimize inference latency. Table \ref{table.comp} compares \textit{HiDP} against different workload partitioning strategies for edge-edge and edge-cloud paradigms. Unlike the \ac{soa} strategies, \textit{HiDP} considers latency and energy reduction through hierarchical partitioning.
\section{HiDP Framework}\label{sec.frm}


\textit{HiDP} is a 2-step workload splitting framework for distributed \ac{dnn} inference across heterogeneous edge clusters. The nodes are connected via wireless networks and can collaborate at run-time to perform inference tasks. The \ac{dnn} inference requests arrive randomly at a local node and the \textit{HiDP} framework performs hierarchical partitioning to achieve low inference latency. As shown in the left side of Figure~\ref{fig:sys}, the framework includes \textit{Application Module}, \textit{Communication Module}, \textit{Global partitioner}, \textit{Local partitioner}, and a \textit{Run-time scheduler}. The right side of the figure shows the scenario of distributed \ac{dnn} inference using \textit{HiDP} framework across a heterogeneous cluster of four edge nodes. The \textit{HiDP} framework receives a \ac{dnn} inference request in the \textit{Application module} of \textit{Device-1} and sends the \ac{dnn} to \textit{Run-time scheduler}. We designed a scheduling policy in the \textit{Run-time scheduler} module that monitors and controls the workload splitting and distribution across the edge cluster. The \textit{Run-time scheduler} gets the status of the cluster-wide node availability and invokes the \textit{Global partitioner} to find the optimal partitioning point. The \textit{Global partitioner} consults a \ac{dse} agent to find the optimal partitioning mode and the feasible partitions. The \textit{Global partitioner} selects model partitioning; then it distributes the workload across the edge cluster via \textit{Communication Module}. The \textit{Communication Module} provides access to send and receive data across the edge cluster. The \textit{Global partitioner} sends the local partition to the \textit{Run-time scheduler} for local execution. The \textit{Run-time scheduler} invokes the \textit{Local partitioner} to find the optimal partitioning point and mode using a \textit{DSE} agent. The \textit{Local partitioner} selects data partitioning and splits the workload following heterogeneity of two CPUs and one GPU. The \textit{Run-time scheduler} gets the local and global results via \textit{Communication Module} and merges all the results. 

\begin{table}[]
\vspace{-12pt}
\caption{Comparison of HiDP with other relevant approaches}
\vspace{-6pt}
\setlength\tabcolsep{4.8pt}
\scalebox{0.88}{
\begin{tabular}{lccccc}
\hline
\multicolumn{1}{|l|}{\textbf{}} & \multicolumn{1}{c|}{\textbf{\begin{tabular}[c]{@{}c@{}}Partition\\ type\end{tabular}}} & \multicolumn{1}{c|}{\textbf{\begin{tabular}[c]{@{}c@{}}Target\\ platform\end{tabular}}} & \multicolumn{1}{c|}{\textbf{\begin{tabular}[c]{@{}c@{}}Global\\ Partitioning\end{tabular}}} & \multicolumn{1}{c|}{\textbf{\begin{tabular}[c]{@{}c@{}}Local\\ Partitioning\end{tabular}}} & \multicolumn{1}{c|}{\textbf{\begin{tabular}[c]{@{}c@{}}heterogeneous\\ block size\end{tabular}}} \\ \hline
\multicolumn{1}{|l|}{\cite{DeepThings}} & \multicolumn{1}{c|}{Data} & \multicolumn{1}{c|}{Edge cluster} & \multicolumn{1}{c|}{\ding{51}} & \multicolumn{1}{c|}{\ding{56}} & \multicolumn{1}{c|}{\ding{56}} \\ \hline
\multicolumn{1}{|l|}{\cite{AutoDice}} & \multicolumn{1}{c|}{Data} & \multicolumn{1}{c|}{Edge cluster} & \multicolumn{1}{c|}{\ding{51}} & \multicolumn{1}{c|}{\ding{56}} & \multicolumn{1}{c|}{\ding{51}} \\ \hline
\multicolumn{1}{|l|}{\cite{omniboost}} & \multicolumn{1}{c|}{Model} & \multicolumn{1}{c|}{Edge-cluster} & \multicolumn{1}{c|}{\ding{51}} & \multicolumn{1}{c|}{\ding{56}} & \multicolumn{1}{c|}{\ding{51}} \\ \hline
\multicolumn{1}{|l|}{\cite{roadrunner}} & \multicolumn{1}{c|}{Model} & \multicolumn{1}{c|}{Edge-cloud} & \multicolumn{1}{c|}{\ding{51}} & \multicolumn{1}{c|}{\ding{56}} & \multicolumn{1}{c|}{\ding{51}} \\ \hline
\multicolumn{1}{|l|}{\cite{disnet}} & \multicolumn{1}{c|}{Hybrid} & \multicolumn{1}{c|}{Edge cluster} & \multicolumn{1}{c|}{\ding{51}} & \multicolumn{1}{c|}{\ding{56}} & \multicolumn{1}{c|}{\ding{51}} \\ \hline
\multicolumn{1}{|l|}{HiDP} & \multicolumn{1}{c|}{Hybrid} & \multicolumn{1}{c|}{Edge cluster} & \multicolumn{1}{c|}{\ding{51}} & \multicolumn{1}{c|}{\ding{51}} & \multicolumn{1}{c|}{\ding{51}} \\ \hline
\end{tabular}
}
\vspace{-18pt}
\label{table.comp}
\end{table}

\noindent\textbf{Platform.} 
The edge nodes are supported by an \ac{os} with relevant libraries and programming framework to run \ac{dnn} inference, allowing inter-node communication, workload scheduling, and application-to-core mapping. The \ac{os} provides interfaces between software-software modules to exchange data, and software-hardware modules to bind applications to selected processors. Unlike \ac{soa} strategies, \textit{HiDP} overtakes the control from default \ac{os} governors and allocates the workload to the desired processing units. The \ac{os} allows run-time monitoring of the board's power consumption through onboard sensors or external power monitoring equipment to measure the energy consumption of a \ac{dnn} inference. 

\noindent\textbf{Workloads.}
We have designed \textit{HiDP} to accept unknown \ac{dnn} inference workload without prior knowledge of the workload arrival time. The target applications are streaming applications that generate video and image data for live processing. These applications generate scenario-based input data with variable input sizes and batches. We consider the modern example scenario of a person bearing different smart gadgets and wearables including a smartwatch, smartphone, smart ring, and augmented reality gear. Manufacturers like Apple and Samsung have produced a series of smart devices that can communicate with each other at run-time sharing notifications and live data for a single user. These devices have diverse \ac{dnn} applications that perform cognitive vision tasks of variable input sizes and data volume using similar \ac{dnn} models depending on the requirements of the application tasks.




\noindent \textbf{System Model.} 
We consider \ac{dnn} model as \ac{dag} since the data flow is sequential without loops and each partition is executed only once. The \ac{dag} nodes represent the \ac{dnn} layers and the edges represent the tensors. The \ac{dnn} is denoted as $\mathcal{D}(\mathcal{L}_{i}) = \{\mathcal{L}1, \mathcal{L}2,..., \mathcal{L}_{i}\}$, where $\mathcal{L}$ represents set of layers that can be partitioned following the model and data partitioning. The layer types include convolution, pooling, flatten, or dense, where each layer can be represented as a vector of kernel size, stride, padding, number of input channels, number of output channels, and height of the input dimensions. 
We denote edge cluster as $\mathcal{N}(\phi_{j}) = \{\phi1, \phi2, ..., \phi_{j}\}$, where $\phi_{j}$ represents the edge node.
For each node, there exist $k$ processors such that $\phi = \{\rho_1, \rho_2,..., \rho_k\}$ where $\rho_k$ represents CPU, GPU, or NPU. We denote the computation frequency of a processor as $f_{k}$ as computation cycles per second.
We define the compute intensity of a \ac{dnn} as $\delta$, representing the average number of compute cycles of a processor to execute 1 flop [cycles/flops].
We define the computation rate [flops/sec] of each processor as the ratio of the computation frequency of the processor to the compute intensity of the \ac{dnn}, such as $\lambda = \frac{f_{k}}{\delta}$ \cite{coedge}.
We denote the communication rate of each processor as a scalar $\mu_k$, representing the \ac{dnn} transmission overhead between two processors for a given time duration $t$. We calculate the local computation-to-communication ratio of a node as:
\begin{equation}
\vspace{-3pt}
\psi\{\lambda, \mu\} =
\left\{
\begin{array}{cccc}
\frac{\lambda_1}{\mu_1}, & \frac{\lambda_2}{\mu_2}, &\cdots&, \frac{\lambda_k}{\mu_k}
\end{array}
\right\}
\label{eqs.loc_com-comp}
\end{equation}
Finally, we calculate the computation rate $\Lambda_j$ of a node $\phi_{j}$ as the sum of computation rates of the available processors:
\begin{equation}
\vspace{-5pt}
\Lambda_{j}(\rho_k) = \sum_{1}^k[\lambda_k]
\label{eqs.com_rate}
\end{equation}
We denote the communication rate of each node as a scalar $\beta_{\phi_j}$ representing the \ac{dnn} data transmission overhead between two nodes for a given time duration $t$ via a wireless network. \textit{HiDP} calculates the communication rate of each node by sending a set of pseudo packets to each node and recording the time taken to get the response. We form a global resource vector $\Psi$ including the global computation-to-communication ratio of all nodes such that:
\begin{equation}
\vspace{-3pt}
\Psi\{\Lambda, \beta\} =
\left\{
\begin{array}{cccc}
\frac{\Lambda_1(\rho_k)}{\beta_{\phi_1}}, & \frac{\Lambda_2(\rho_k)}{\beta_{\phi_2}}, & \cdots & \frac{\Lambda_j(\rho_k)}{\beta_{\phi_j}}
\end{array}
\right\}
\label{eqs.glb_com-comp}
\end{equation}

\textit{HiDP} formulates an availability vector $\mathcal{A}(\mathcal{N}_{\phi})$ based on the communication rate $\beta^{(t)}_{\phi}$ such that:
\begin{equation}
\mathcal{A}(\mathcal{N}_{\phi}) = \{\alpha_1, \alpha_2, \cdots, \alpha_j\}
\quad \text{,} \quad
\alpha_j = 
\begin{cases} 
1 & \text{available}, \\
0 & \text{unavailable}
\end{cases}
\end{equation}
For workload partitioning, \textit{HiDP} decides between the partitioning modes and distributes the workload to the resources. For model partitioning, we denote the width of a layer block as $\omega_i$ and calculate the total computation time as:
\begin{equation}
\vspace{-3pt}
\Theta_{\omega} = 
\left\{
\begin{array}{cccc}
\gamma \cdot \omega
\end{array}
\right\}
\quad \text{,} \quad
\gamma = 
\begin{cases} 
\Psi & \text{for global partitioning} \\
\psi & \text{for local partitioning}
\end{cases}
\label{eqs.model_part}
\vspace{-3pt}
\end{equation}
Similarly, \textit{HiDP} explores the number of parallel submodels $\sigma$ for data partitioning to calculate the total computation time as:  
\begin{equation}
\vspace{-3pt}
\Theta_{\sigma} = 
\left\{
\begin{array}{cccc}
\gamma \cdot \sigma
\end{array}
\right\}
\quad \text{,} \quad
\gamma = 
\begin{cases} 
\Psi & \text{for global partitioning} \\
\psi & \text{for local partitioning}
\end{cases}
\label{eqs.data_part}
\end{equation}
\textit{HiDP} calculates the total computation time $\Theta$ of both model and data partitioning modes and selects the faster strategy.


\noindent \textbf{Scheduling Algorithm.}
Algorithm 1 shows the high-level workflow of \textit{HiDP} framework to perform distributed inference. \textit{HiDP} assigns leader status ($\phi_1^*$) to the node that receives new inference request $\mathcal{D_L}$ (Lines 1--2). The leader node checks the availability status ($\mathcal{A}(\mathcal{N}_\phi)$) of the cluster (Line-3) and finds the optimal partitioning mode using \ac{dp} algorithm (Lines 4--6). We used a standard subset sum algorithm for an efficient recursive search with time complexity $O(n * m)$, where $n$ represents the number of \ac{dnn} \textit{blocks} and $m$ represents the number of available nodes.
The algorithm starts with the largest possible block sizes following the resource heterogeneity to calculate the inference latency and back-propagates block by block to converge to minimum latency. We used the same algorithm to explore global and local partitioning points because the function arguments are essentially the same in either case including the \ac{dnn} and the computation-communication ratio. 
The leader node partitions the workload and distributes the partitions to the global nodes (Line 7). The leader node finds out the optimal partitioning mode and the partitioning point for the local partition (Lines 8--10). The leader node executes the local workload on its local processors and gathers the global results (Lines 11--12). The leader node merges the final results and reports the prediction to the \ac{dnn} application (Line 13).

\begin{figure}
    \centering
    \includegraphics[width=0.9\columnwidth]{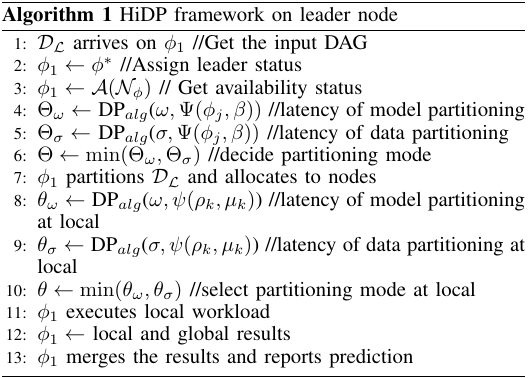}
    \vspace{-15pt}
    \label{fig:alg}
\end{figure}

\begin{figure}
    \centering
    \vspace{-6pt}
    \includegraphics[width=0.45\textwidth]{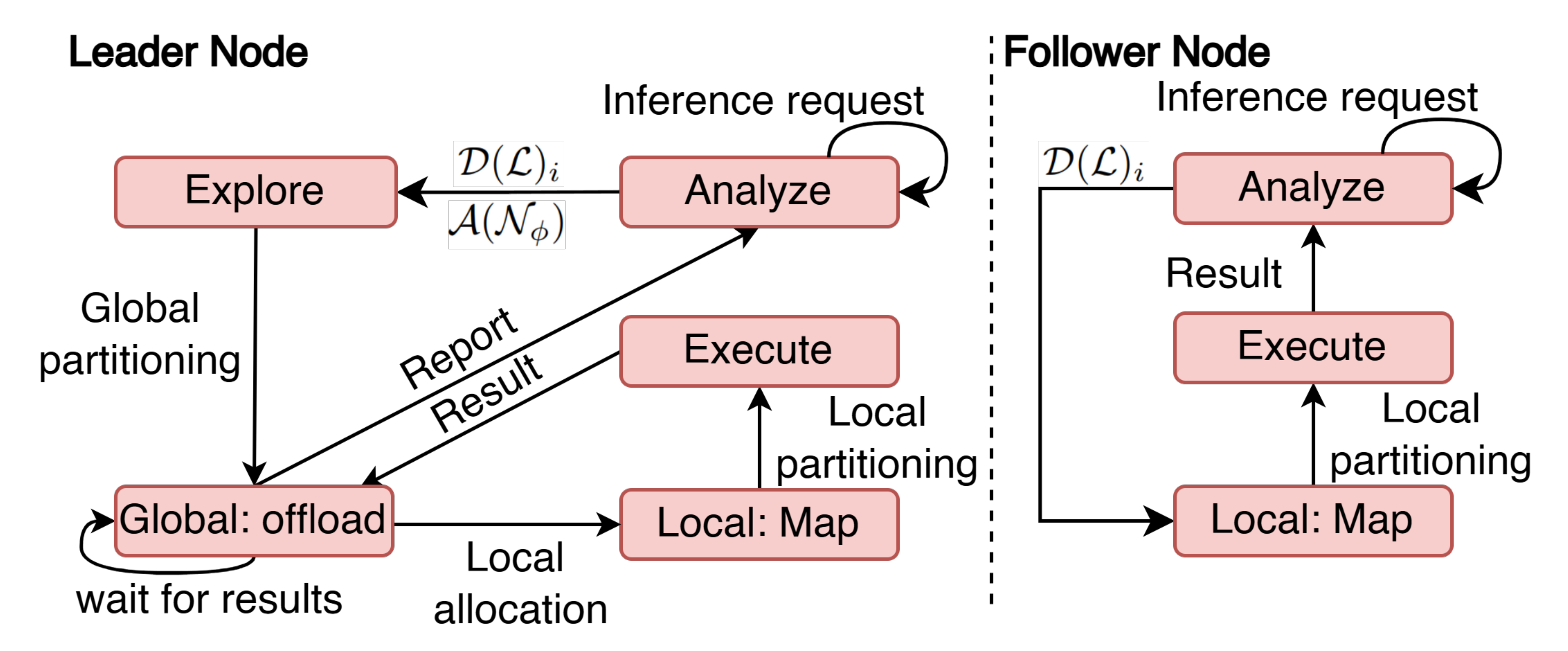}
    \vspace{-9pt}
    \caption{Workflow of the leader and follower nodes in the \textit{HiDP} controller}
    \vspace{-15pt}
    \label{fig:fsm}
\end{figure}

\noindent\textbf{Run-time Scheduler.}
We have designed the scheduling policy of the \textit{Run-time Scheduler} as a \ac{fsm} including \textit{Analyze, Explore, Offload, Map,} and \textit{Execute} states as shown in Figure \ref{fig:fsm}. The scheduling policy is the implementation on the method explained in Algorithm 1.
The scheduling policy is different for the leader and follower nodes.

\noindent \textbf{\textit{Leader Node.}} In the \textit{Analyze} state, the controller waits for an inference request from a \ac{dnn} application in the \textit{Application module}. When a request is triggered, the controller checks the availability status of the cluster nodes by sending and receiving a status packet to the nodes via \textit{Communication module} and transitions to the \textit{Explore} state. In the \textit{Explore} state, the controller refers to the global \ac{dse} agent to find the optimal partitioning point of the given \ac{dnn} workload for global distribution.
After finding the optimal partitioning point, the controller transitions to the \textit{Global: offload} state. The controller offloads the workload to the available nodes using the \textit{Communication module} and transitions to the \textit{Local: Map} state for local execution of the allocated workload. Here, the controller refers to the local \ac{dse} agent to figure out the local partitioning of the workload following the available processing units.
After converging to the optimal partitioning point, the controller transitions to the \textit{Execute} state. In this state, the controller executes the workload while sharing intermediate data with the available nodes for parallel or sequential execution, depending on the partitioning mode.
After successful execution, the controller gathers the results and transitions back to the \textit{Global: offload} state for final merging and reporting of the results. After merging the results from local execution and cluster nodes, the controller transitions to the \textit{Analyze} state and waits for the next inference request. 

\noindent \textbf{\textit{Follower Nodes.}} For the follower nodes, the state machine is much simpler, such that (i) the node receives the distributed workload from the leader node in the \textit{Analyze} state, (ii) executes it after local partitioning in the \textit{Local: Map} and the \textit{Execute} state, and (iii) report back the results to the leader.

\section{Experimental Evaluation} \label{sec.exp}


\subsection{Experimental Setup}
\noindent\textbf{Evaluation Platform.}
We design a collaborative heterogeneous edge cluster, comprising commercial edge platforms (shown in Table \ref{tab.specs}), and deploy the \textit{HiDP} strategy on the defined platform. We monitor the run-time power consumption of the Jetson boards using the on-board power sensors. We use the external shunt resistor to monitor the power consumption of the Raspberry Pi boards. 

\noindent\textbf{Workloads.} We evaluated our framework over ResNet152, EfficientNetB0, VGG-19, and InceptionNetV3, which are widely used \ac{dnn} models in mobile vision applications.
We implemented these models using the TensorFlow library \cite{tensorflow}, with input image sizes of 224x224, and 299x299. We enhanced the top layer of these models to accept variable input sizes to enable data partitioning. 

\begin{table}[t]
\vspace{-3pt}
\centering
\caption{Technical specifications of the evaluation setup}
\vspace{-3pt}
\scalebox{0.82}{
\begin{tabular}{lcccccc}
\hline
\textbf{Device} & \textbf{CPU} & \textbf{GPU} & \textbf{DRAM}\\
\hline
Jetson Orin NX & 8x ARM Cortex-A78 & 1024-core Ampere & 8 GB\\
\rowcolor{gray!25}
Jetson TX2 & 2x Denver-2, 4x ARM Cortex-A57 & 256-core Pascal & 8 GB\\
\rowcolor{gray!25}
Jetson Nano & 4x ARM Cortex-A57 & 128-core Maxwell & 4 GB\\
Raspberry Pi 5 & 2x ARM Cortex-A76 & VideoCore VII & 4 GB\\
\rowcolor{gray!25}
Raspberry Pi 4 & 2x ARM Cortex-A72 & VideoCore VI & 4 GB\\
\hline
\end{tabular}
}
\vspace{-12pt}
\label{tab.specs}
\end{table}

\noindent\textbf{Middleware.}
We implemented \textit{HiDP} as a middleware in Python with a source code of 400 lines. Each device hosts Linux 18.04 \ac{os} to provide software and hardware interfaces. We used CGroup libraries to bind the workload to the desired number of CPU cores. For GPU implementation, the TensorFlow backend used CUDA libraries for NVIDIA boards and OpenGL for Raspberry Pi boards. 
All the devices are connected over 80 MBps wireless control through POSIX-based client-server architecture. We used multi-threaded server operations for the leader nodes to communicate with the available nodes dynamically. The overhead of using \ac{dp} algorithm-based exploration including both global and local partitioning is 15ms on average for our experimentation.

\noindent\textbf{Comparison w.r.t. state-of-the-art approaches.} For evaluation, we considered three different state-of-the-art partitioning strategies -- data\cite{MoDNN}, model \cite{omniboost}, and hybrid \cite{disnet}. 
\textit{MoDNN} \cite{MoDNN} partitions and distributes the input data proportionally among the available edge nodes. 
We implemented \textit{MoDNN} using the data partitioning module of \textit{HiDP} framework. 
OmniBoost \cite{omniboost} determines the optimal partitioning point using the Monte-Carlo search tree and pipelines the \ac{dnn} inference over both CPU and GPU. We implemented the throughput estimator of \textit{Omniboost} using Gymnasium library \cite{gym} and trained it on our target workloads.
\textit{DisNet} \cite{disnet} uses heuristic-based \ac{dnn} partitioning and distribution by jointly considering data and model partitioning. We used the data and model partitioning algorithm of \textit{HiDP} to implement \textit{DisNet}.

\vspace{-6pt}
\subsection{Experimental Results}
\begin{figure}
    \centering
    \includegraphics[width=0.48\textwidth]{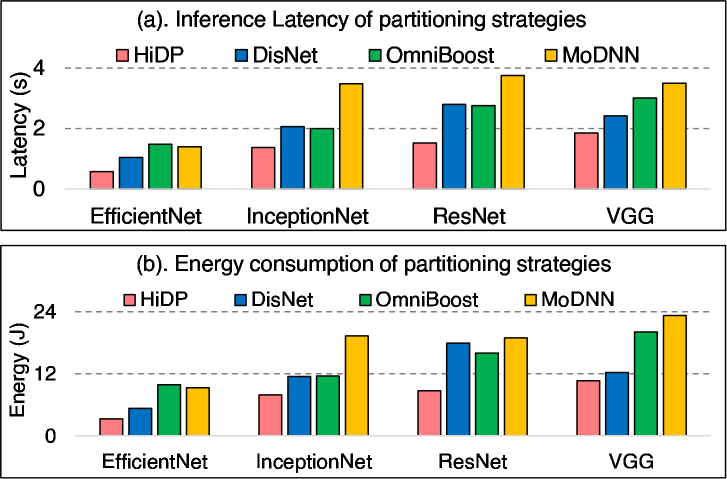}
    \vspace{-9pt}
    \caption{Inference (a). latency and (b). energy consumption of different strategies for targeted \ac{dnn} workloads.}
    \vspace{-9pt}
    \label{fig:latency}
\end{figure}

For our experiments, we consider inference latency, throughput, energy consumption (based on run-time power monitoring), and accuracy as evaluation metrics.
Figure \ref{fig:latency}~(a) shows the inference latency of EfficientNetB0, InceptionNetV3, ResNet152, and VGG-19 using different strategies.
Our proposed \textit{HiDP} strategy has the lowest inference latency for all the workloads, in comparison with other relevant strategies, achieving upto 61\%, 61\%, 59\%, and 49\% lower latency for EfficientNet, InceptionNet, ResNet, and VGG, respectively. 
The hierarchical partitioning strategy of \textit{HiDP} jointly optimizes both global-level \ac{dnn} block creation and workload distribution, followed by local-level fine-grained partitioning and workload scheduling. This results in significantly lower latency compared to other distributed inference strategies that are exclusively confined to global partitioning.
On average, \textit{HiDP} has 37\%, 44\%, and 56\% lower latency than \textit{DisNet, OmniBoost,} and \textit{MoDNN} strategies, respectively. Figure \ref{fig:latency} (b) shows the energy consumption of different partitioning strategies for all the workloads. The lowest inference latency of \textit{HiDP} strategy also reflects in the lowest energy consumption for all the workloads. \textit{HiDP} consumes upto 67\%, 59\%, 54\%, and 54\% lower energy for EfficientNet, inceptionNet, ResNet, and VGG against relevant strategies. On average \textit{HiDP} consumes  33\%, 48\%, and 58\% lower energy than DisNet, OmniBoost, and MoDNN, respectively.

\begin{figure}
    \centering
    \includegraphics[width=0.48\textwidth]{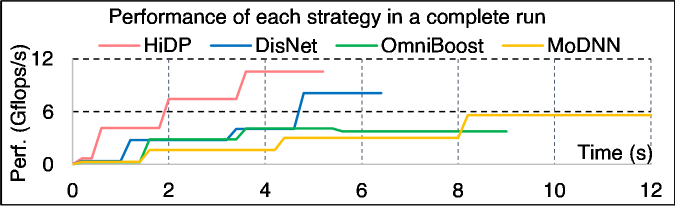}
    \vspace{-9pt}
    \caption{Performance (Gigaflops/s) of each strategy while concurrently running targeted \ac{dnn} workloads.}
     \vspace{-18pt}
    \label{fig:perf}
\end{figure}

We evaluate the adaptability of different strategies under varying workload dynamics. We created a dynamic workload with successive run-time inference requests for every 0.5s, in the order of EfficientNetB0, InceptionNetV3, ResNet152, and VGG-19. This creates a progressively increasing workload such that at t=1.5s, all four \acp{dnn} are running concurrently on the edge cluster.
Figure \ref{fig:perf} shows the performance (Gigaflops per second) of each strategy while running different \ac{dnn} models. It can be noticed that \textit{HiDP} delivers the highest performance throughout the execution cycle. Lower latency achieved by \textit{HiDP} frees up the resources of edge nodes, which can be efficiently used to service subsequent inference requests. \textit{HiDP} completes the inference of all the models within 5s in total, achieving 39\%, 54\%, and 56\% higher performance than \textit{DisNet, OmniBoost,} and \textit{MoDNN}, respectively. Higher per-inference latency of other strategies keeps the worker edge nodes busy for a longer duration, affecting the overall performance and throughput. We also evaluate the throughput (number of inferences per 100s) of different strategies over 8 different mixes of \ac{dnn} inference requests.
We created \textit{Mix 1-4} and \textit{Mix 5-8} with two and three different \ac{dnn} models from the target workloads, respectively. 
\textit{HiDP} achieves significantly higher throughput (upto 150\% in \textit{Mix-2} and 56\% on average) compared to other strategies across all the workload mixes. \textit{HiDP} dynamically selects data/model partitioning based on \ac{dnn} characteristics, flexibly achieving low latency inference across different workload mixes. This is reflected in higher throughput achieved by \textit{HiDP} in comparison with other strategies that are confined to specific partitioning configurations.
\begin{figure}
    \centering
    \includegraphics[width=0.45\textwidth]{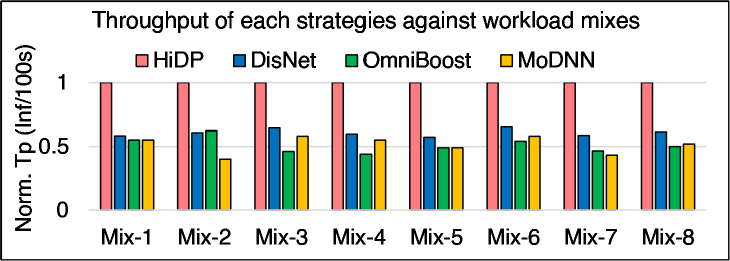}
    \vspace{-3pt}
    \caption{Throughput of different strategies while running different combinations of targeted \ac{dnn} workloads concurrently.}
    \vspace{-6pt}
    \label{fig:tp}
\end{figure}

\begin{figure}
    \centering
    \vspace{-3pt}
    \includegraphics[width=0.45\textwidth]{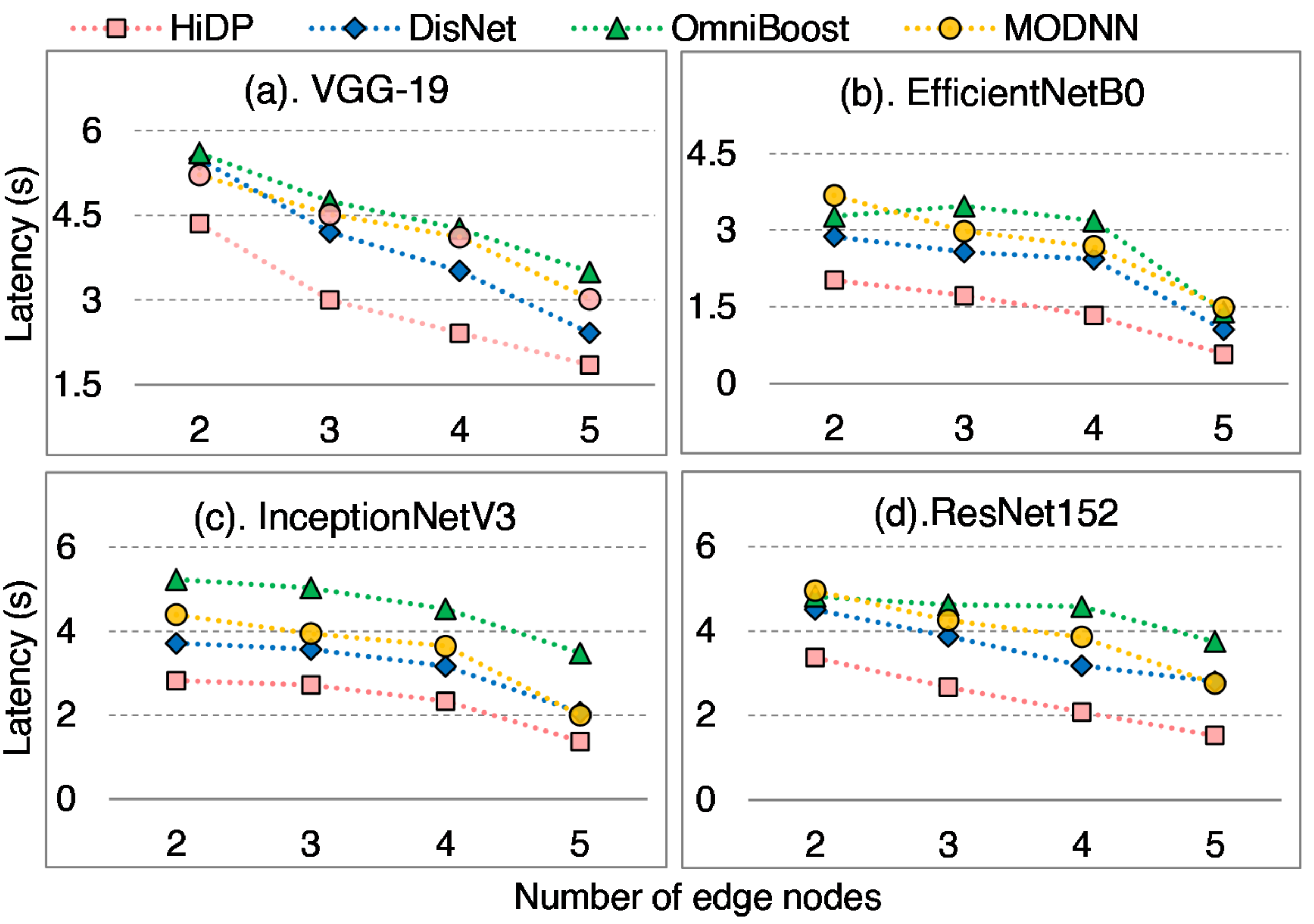}
    \vspace{-9pt}
    \caption{Inference latency with varying number of worker edge nodes.}
    \vspace{-9pt}
    \label{fig:work}
\end{figure}
\vspace{-4pt}

We evaluate the adaptability of different strategies with varying numbers of edge devices within the cluster. Figure \ref{fig:work} shows inference latency of all the \ac{dnn} workloads using different strategies with 2-5 edge nodes. For all the workloads and varying numbers of edge nodes, \textit{HiDP} has the lowest inference latency. It should be noted that the latency gap between \textit{HiDP} and other strategies becomes prominent with a decreasing number of worker edge nodes. This can be attributed to \textit{HiDP}'s efficient utilization of local edge node resources, while the other global strategies are affected by a lower number of edge devices. On average \textit{HiDP} achieves 30\%, 46\%, and 38\% lower latency than DisNet, OmniBoost, and MoDNN, respectively. For VGG-19, EfficientNetB0, ResNet-152, and InceptionNet-V3, \textit{HiDP} has an average Top-1\% accuracy of 75.3\%, 77.1\%, 78.6\%, and 80.9\% and Top-5 \% accuracy of 89.7\%, 92.25\%, 92.7\%, and 92.5\%, respectively. Both the Top-1\%, and Top-5\% accuracies of \textit{HiDP} are the same as DisNet, OmniBoost, and MoDNN, demonstrating robust intermediate data sharing while enforcing \ac{dnn} partitioning.  

\section{Conclusion}\label{sec.conc} \vspace{-3pt}
We present \textit{HiDP} framework for low latency \ac{dnn} inference on distributed edge platforms. Our approach hierarchically partitions the \ac{dnn} inference workload at a global level, followed by optimized partitioning and scheduling on a local heterogeneous edge node. We evaluated \textit{HiDP} on four Jetson platforms and two Raspberry Pi platforms achieving latency and energy improvements of 38\%, and 46\% against \ac{soa} strategies, respectively. We consider energy-efficient distributed inference for future work.


\bibliographystyle{IEEEtran}
\bibliography{references}

\end{document}